\documentclass[10pt,journal,compsoc]{IEEEtran}

\ifCLASSOPTIONcompsoc
  \usepackage[nocompress]{cite}
\else
  \usepackage{cite}
\fi

\ifCLASSINFOpdf
  \usepackage[pdftex]{graphicx}
\else
  \usepackage[dvips]{graphicx}
\fi

\usepackage{flushend}

\ifCLASSOPTIONcompsoc
  \usepackage[caption=false,font=footnotesize,labelfont=sf,textfont=sf]{subfig}
\else
  \usepackage[caption=false,font=footnotesize]{subfig}
\fi

\ifCLASSOPTIONcaptionsoff
  \usepackage[nomarkers]{endfloat}
 \let\MYoriglatexcaption\caption
 \renewcommand{\caption}[2][\relax]{\MYoriglatexcaption[#2]{#2}}
\fi

\usepackage{amsmath}
\usepackage{amssymb}
\usepackage{empheq}

\usepackage{tikz}
\usepackage{pgfplots}
\usetikzlibrary{shapes,arrows,calc,patterns,decorations.pathmorphing,decorations.markings,circuits.ee.IEC}
\usepackage{circuitikz}
\usetikzlibrary{calc}

\usepackage[printonlyused,smaller]{acronym}

\acrodef{AAC}{Advanced Audio Coding}
\acrodef{AC}{alternate current}
\acrodef{ADC}{Analog Digital Converter}
\acrodef{AGU}{Address Generating Unit}
\acrodef{ALU}{Algorithmic Logical Unit}
\acrodef{AVM}{Astute Virtual Machine}
\acrodef{ASIC}{Application-Specific Integrated Circuit}
\acrodef{BACPAC}{Berkeley Advanced Chip Performance Calculator}
\acrodef{BEEBS}{Bristol Energy Efficiency Benchmark Suite}
\acrodef{BGA}{ball grid array}
\acrodef{BSIM}{Berkeley Short-channel {\smaller IGFET} Model}
\acrodef{BTBT}{band-to-band tunneling}
\acrodef{CAD}{computer-aided design}
\acrodef{CISC}{Complex Instruction Set Computer}
\acrodef{CMOS}{complementary metal-oxide-semiconductor}
\acrodef{CFD}{computational fluid dynamics}
\acrodef{CPU}{Central Processing Unit}
\acrodef{CPI}{clock-cycles per instruction}
\acrodef{DC}{direct current}
\acrodef{DDIO}{Data Direct I/O}
\acrodef{DFG}{Data Flow Graph}
\acrodef{DIBL}{drain-induced barrier lowering}
\acrodef{DLP}{Data Level Parallelism}
\acrodef{DMA}{Direct Memory Access}
\acrodef{DRAM}{Dynamic Random-Access Memory}
\acrodef{DSP}{Digital Signal Processor}
\acrodef{DTM}{Dynamic Thermal Management}
\acrodef{DVS}{Dynamic Voltage Scaling}
\acrodef{DFS}{Dynamic Frequency Scaling}
\acrodef{DVFS}{Dynamic Voltage and Frequency Scaling}
\acrodef{DPM}{Dynamic Power Management}
\acrodef{EDD}{Energy-Delay Diagram}
\acrodef{EMA}{exponential moving average}
\acrodef{EEPROM}{Electrically Erasable Programmable Read-Only Memory}
\acrodef{EER}{Energy Efficiency Rating}
\acrodef{FFT}{Fast Fourier Transformation}
\acrodef{FPA}{Floating Point Adder}
\acrodef{FPU}{Floating Point Unit}
\acrodef{FPM}{Floating Point Multiplier}
\acrodef{FMA}{fused multiply?add}
\acrodef{FSM}{Finite State Machine}
\acrodef{GIDL}{gate-induced drain leakage}
\acrodef{GPU}{Graphics Processing Unit}
\acrodef{GPS}{Global Positioning System}
\acrodef{GSM}{Global System for Mobile Communications}
\acrodef{GUI}{graphical user interface}
\acrodef{HC}{Hardware Counter}
\acrodef{HCI}{human-computer interaction}
\acrodef{HDL}{Hardware Description Language}
\acrodef{HPCo}[HPC]{hardware performance counter}
\acrodef{HPC}{High Performance Computing}
\acrodef{IC}{Integrated Circuit}
\acrodef{iff}{if and only if}
\acrodef{IR}{infrared}
\acrodef{ILP}{integer linear programming}
\acrodef{IP}{integer programming}
\acrodef{IoT}{Internet of Things}
\acrodef{I2C}[I$^2$C]{Inter-Integrated Circuit}
\acrodef{IQR}{interquartile range}
\acrodef{ITRS}{In\-ter\-na\-tion\-al Tech\-nolo\-gy Road\-map for Semi\-con\-duc\-tors}
\acrodef{ILP}{Instruction Level Parallelism}
\acrodef{I/O}{input/output}
\acrodef{ISA}{Instruction Set Architecture}
\acrodef{FIR}{finite impulse response}
\acrodef{JIT}{Just-In-Time}
\acrodef{JNI}{Java Native Interface}
\acrodef{LPDDR}{low power DRAM}
\acrodef{LCD}{liquid crystal display}
\acrodef{MAD}{median absolute deviation}
\acrodef{MAE}{maximum absolute error}
\acrodef{MP3}{{\small MPEG-1} or {\small MPEG-2} Audio Layer {\small III}}
\acrodef{MIPJ}{millions-of-instructions-per-joule}
\acrodef{MOSFET}{metal-oxide semiconductor field-effect transistor}
\acrodef{MPSoC}{multiprocessor System-on-Chip}
\acrodef{MSE}{mean squared error}
\acrodef{MTTF}{Mean Time To Failure}
\acrodef{MIPS}{Microprocessor without Interlocked Pipeline Stages}
\acrodef{NIC}{Network Interface Card}
\acrodef{NIST}{National Institute of Standards and Technology}
\acrodef{NDK}{Native Development Kit}
\acrodef{NSVS}{nonsupply-voltage-scaling}
\acrodef{NTP}{Normal Temperature and Pressure}
\acrodef{PoP}{Package-on-Package}
\acrodef{PCB}{printed circuit board}
\acrodef{PID}{proportional-integral-derivative}
\acrodef{OOE}{out-of-order execution}
\acrodef{OS}{Operating System}
\acrodef{QME}{quantum mechanical effects}
\acrodef{RAM}{Random Access Memory}
\acrodef{RISC}{Reduced Instruction Set Computing}
\acrodef{ROHC}{Robust Header Compression}
\acrodef{RMS}{Root Mean Square}
\acrodef{rpm}{revolutions per minute}
\acrodef{RTL}{Register Transfer Language}
\acrodef{RC}{resistor/capacitor}
\acrodef{RMSE}{root-mean-square error}
\acrodef{SISD}{Single Instruction Single Data}
\acrodef{SIMD}{Single Instruction Multiple Data}
\acrodef{SMA}{simple moving average}
\acrodef{SMD}{surface mount device}
\acrodef{SoC}{Systems-on-Chip}
\acrodef{SGLP}{Super-graph Level Parallelism}
\acrodef{SLP}{Super-word Level Parallelism}
\acrodef{SMT}{simultaneous multi\-thread\-ing}
\acrodef{SPM}{Scratch-Pad Memory}
\acrodef{SVM}{State Vector Machine}
\acrodef{SRAM}{Static Random-access Memory}
\acrodef{SDRAM}{synchronous dynamic random access memory}
\acrodef{STP}{standard temperature and pressure}
\acrodef{TCP}{Transport Control Protocol}
\acrodef{TDP}{thermal design power}
\acrodef{TCT}{Task Completion Time}
\acrodef{TLB}{Translation Look-aside Buffer}
\acrodef{TLP}{Thread Level Parallelism}
\acrodef{TP}{Travaux Pratiques}
\acrodef{TMU}{Thermal Management Unit}
\acrodef{TTA}{Transport-Triggered Architecture}
\acrodef{UMTS}{Universal Mobile Telecommunications System}
\acrodef{VA}{volt-ampere}
\acrodef{VC}{Virtual Channel}
\acrodef{VM}{Virtual Machine}
\acrodef{VLSI}{Very-Large-Scale Integration}
\acrodef{VHDL}{VHSIC Hardware Description Language}
\acrodef{VLIW}{Very Long Instruction Word}
\acrodef{VM}{Virtual Machine}
\acrodef{WFL}{Weber-Fechner law}
\acrodef{WiFi}{Wireless-Fidelity}
\acrodef{WLAN}{Wireless Local Area Network}
\acrodef{WSN}{Wireless Sensor Network}

\newcommand{\beebs}{{\small BEEBS}}

\newcommand{\ea}{\emph{et al.}}

\newcommand{\efcr}{Energy/Frequency Convexity Rule}

\newcommand{\Pcpu}{P_\mathrm{\smaller cpu}}
\newcommand{\Pcpui}{P_{\mathrm{\smaller cpu},i}}
\newcommand{\Pdrop}{P_\mathrm{\smaller drop}}
\newcommand{\Pdropi}{P_{\mathrm{\smaller drop},i}}

\newcommand{\tx}{\Delta t}

\newcommand{\Ps}{P_{\mathrm{sys}}}
\newcommand{\Psii}{P_{\mathrm{sys},i}}
\newcommand{\Dti}{\Delta t_i}

\newcommand{\Es}{E_{\mathrm{sys}}}
\newcommand{\Esi}{E_{\mathrm{sys},i}}

\newcommand{\En}{E_\mathrm{n}}

\newcommand{\Pb}{P_{\mathrm{back}}}

\newcommand{\cb}{cc_{\mathrm{b}}}
\newcommand{\cbi}{cc_{\mathrm{b},i}}

\newcommand{\ck}{f_{\mathrm{k}}}

\newcommand{\cki}{f_{\mathrm{k},i}}

\newcommand{\fo}{f_\mathrm{opt}}

\newcommand{\fc}{f_{\mathrm{cpu}}}
\newcommand{\Fc}{\mathcal{F}_{\mathrm{cpu}}}
\newcommand{\Fe}{\mathcal{F}_{\mathrm{epx}}}

\newcommand{\Pback}{P_{\mathrm{back}}}

\newcommand{\dd}{\mathrm{d}}

\newcommand{\odroidxue}{{\smaller ODROID XU+E}}

\newcommand{\maxx}{\mathrm{max}}
\newcommand{\minn}{\mathrm{min}}

\newcommand{\EnA}{\En^\mathrm{A}}
\newcommand{\EnB}{\En^\mathrm{B}}

\begin{document}

\title{Parameter Sensitivity Analysis of the\\Energy/Frequency Convexity Rule\\for Nanometer-scale Application Processors}

\author{Karel~De\,Vogeleer,
        Gerard~Memmi,
        and~Pierre~Jouvelot
\IEEEcompsocitemizethanks{\IEEEcompsocthanksitem K. De\,Vogeleer and Gerard Memmi are with TELECOM ParisTech -- INFRES -- CNRS LTCI - UMR 5141 -- Paris, France,\protect\\
Email: \{karel.devogeleer,gerard.memmi\}@telecom-paristech.fr
\IEEEcompsocthanksitem Pierre Jouvelot is with MINES ParisTech, PSL Research University, France. Email: pierre.jouvelot@mines-paristech.fr}
}

\IEEEtitleabstractindextext{%
\begin{abstract}
Both theoretical and experimental evidence are presented in this work in order to validate the existence of an Energy/Frequency Convexity Rule, which relates energy consumption and microprocessor frequency for nanometer-scale microprocessors.
Data gathered during several month-long experimental acquisition campaigns, supported by several independent publications, suggest that energy consumed is indeed depending on the microprocessor's clock frequency, and, more interestingly, the curve exhibits a clear minimum over the processor's frequency range.
An analytical model for this behavior is presented and motivated, which fits well with the experimental data.
A parameter sensitivity analysis shows how parameters affect the energy minimum in the clock frequency space.
The conditions are discussed under which this convexity rule can be exploited, and when other methods are more effective, with the aim of improving the computer system's energy management efficiency.
We show that the power requirements of the computer system, besides the microprocessor, and the overhead affect the location of the energy minimum the most.
The sensitivity analysis of the \efcr\ puts forward a number of simple guidelines especially for by low-power systems, such as battery-powered and embedded systems, and less likely by high-performance computer systems.
\end{abstract}

}

\maketitle

\IEEEdisplaynontitleabstractindextext

\IEEEpeerreviewmaketitle

\ifCLASSOPTIONcompsoc
\IEEEraisesectionheading{\section{Introduction}\label{sec:introduction}}
\else
\section{Introduction}
\label{sec:introduction}
\fi

\IEEEPARstart{T}{he} execution time characteristics and power requirements of a code sequence are the main drivers that define its final energy consumption.
This is a direct result of the definition of electrical energy consumption: the integral of electrical power over time.
The execution time is influenced by the type and the amount of operations contained by the code sequence of concern.
For example register-based operations will require less energy to execute compared to external memory-based instructions.
As such, each functional unit within a microprocessor and, more generally, each component of the computer system have their own respective power and execution time profiles.
As a result, every code sequence has different power and execution time demands.
For example, Carroll and Heiser~\cite{Carroll:2010:APC:1855840.1855861} showed that, for an embedded system running \texttt{equake}, \texttt{vpr}, and \texttt{gzip} from the {\small SPEC CPU2000} benchmark suite, the microprocessor energy consumption exceeds the {\small RAM} memory consumption, whereas \texttt{crafty} and \texttt{mcf} from the same suite showed to be straining more energy from the device {\small RAM} memory.

A property of a code sequence's energy consumption is that, under certain assumptions, it shows convex properties, which is henceforth referred to as the \efcr~\cite{paper:karel:warsaw}.
The rule states that there exists an optimum clock frequency for the execution of each sequence of code that minimizes the energy consumption of that code sequence.
Under certain conditions this optimal clock frequency, minimizing energy consumption, lies between the minimum and maximum clock frequency.
The existence of a minimum energy point results from the behavior of the microprocessor's power and the execution time w.r.t. the clock frequency.
The microprocessor's power increases about linearly with clock frequency, meaning that more energy is consumed when the microprocessor's speed is increased.
On the other hand, the slower the clock frequency, the longer execution time will increase the energy expenditure.
As will be shown, running at the optimal clock frequency is a trade-off between performance, in terms of execution time, and energy savings.
For applications requiring human interaction, it has been shown that the clock frequency can be scaled down considerably without affecting user's experience~\cite{DBLP:conf/iiswc/SeekerPLF14}.
In this paper, experimental evidence is presented, supported by several independent publications, for the existence of an \efcr\ that relates energy consumption and microprocessor clock frequency on mobile devices.
This convexity property seems to ensure the existence of an optimal frequency where energy consumption is minimal.
This existence claim is based on both theoretical and practical evidence on a \ac{SoC}.
Data gathered via acquisition campaigns on multiple platforms suggest that the energy consumed per input element is strongly correlated with microprocessor clock frequency and, more interestingly, that the corresponding curve exhibits a clear minimum over a frequency window specific to the computer system.
An analytical model of this behavior is also motivated, which fits well with the experimental data.
A parameter sensitivity analysis is carried out to assess the influence of the parameters on the optimal frequency minimizing energy consumption.
This optimal frequency is shown to increase when the power requirements of the computer system, excluding the microprocessor's, increase.
Clock cycles lost for routine maintenance of the system also force the optimal frequency up.
The optimal frequency as derived from the theoretical framework is, however, independent of the number of instructions to be executed.

In addition to a deeper theoretical and practical understanding of a microprocessor's energy consumption and the \efcr, this paper offers a new, in-depth, parameter sensitivity analysis compared to what was presented in De~Vogeleer~\ea\cite{paper:karel:warsaw}.
The main contributions of this paper are thus:
\begin{itemize}
	\item a theoretical framework for the \efcr;
	\item a sensitivity analysis of the \efcr\ to estimate the impact of multiple input parameters;
	\item an analysis of the \efcr\ under special conditions, such as, \acf{OOE} and absence of slack time;
	\item supportive experimental data and a comprehensive survey of the state of the art.
\end{itemize}

The rest of the paper is organized as follows.
Section~\ref{sec:single_microprocessor_convexity_model} elaborates the \efcr.
Followed by the presentation of experimental results in Section~\ref{chapter:2b:sec:experimental:results}, a parameter sensitivity analysis is carried out in Section~\ref{sec:sensitivity_of_the_convexity_model}.
An overview of the related work is presented in the state-of-the-art Section~\ref{sec:state_of_the_art}.
Finally, Section~\label{sec:chapter:2b:conclusion} lists the main conclusions drawn from our analysis supporting a better usage of the energy especially for embedded systems.

\section{Single-Core Convexity Model}
\label{sec:single_microprocessor_convexity_model}

The energy consumption of a computer system comprising a microprocessor, and possibly other components, over a time interval $\Delta t$, is equal to the integral of its system's power usage over time:
\begin{equation}
 \Es (\Delta t) = \int_0^{\Delta t} \Ps(t)~ \dd t = \int_0^{\Delta t} I(t)\cdot V(t)~\dd t \label{eq:efcr:integral}.
\end{equation}
If the power is considered constant, the integral is equivalent to the product of the power consumption and the timespan of interest.
$V(t)$ can often be considered constant by design; for example, portable devices such as smartphones are supplied by 3.7\,V lithium-ion batteries, and microprocessors operate at very specific voltage levels.
The current's time-dependent variance depends on the context, its history and the state of the microprocessor.
However, at the time frame of an instruction execution, henceforth referred to as a \emph{time quanta}, the energy consumption can be deemed quasi constant.
Following this definition, the parameters that define the energy consumption during a time quanta are also constant.
As such, similar to the rationale behind the \emph{Riemann sum}, the total energy consumption of a code sequence can be thought of as the sum of the energy consumption during each time quanta $\Delta t$:
\begin{equation}
 \Es = \sum_{i = 1}^n \Esi = \sum_{i = 1}^n \Psii \cdot \Dti,\label{eq:efcr:product}
\end{equation}
where $n$ is the number of time quanta.
$\Dti$ is the time frame over which $\Psii$ is constant.
$\Dti$ could be the length of one instruction execution or, when the power variance is negligibly small, $\Dti$ can be the length of an arbitrary-sized code sequence.
One has $\Delta t = \sum_{i=0}^n \Dti$.

The models for the power and execution time are developed separately in the next two subsections .
A more profound expound of the models can be found in De\,Vogeleer~\cite{thesis:kdv:2015}.

\subsection{Power Model}

A computer system's power usage $\Ps$ is the sum of three power components:
\begin{enumerate}
 \item $\Pcpu$, the microprocessor's power,
 \item $\Pdrop$, the system's power usage that is dependent or controllable by the microprocessor, and
 \item $\Pback$, the system's power that is independent of the microprocessor.
\end{enumerate}
$\Pdrop$ can be due to components that are put to sleep when the microprocessor doesn't need their functionality, e.g., audio codecs, camera circuits, or the radio interface.
$\Pback$ constitutes components that require power independent from what the microprocessor is doing, e.g., memory refreshing in \ac{SDRAM}.
$\Pback$ is however controllable.
It is noted that the display of a hand-held device falls also under $\Pback$ as it is active when the user requires interaction with the device, not necessarily when the microprocessor is active.


For the formulation of the microprocessor's power $\Pcpu$, we combined the well know expression for an electronic circuit's power dissipation $\frac{1}{2}\alpha V^2 f$~\cite{Weste:1985:PCV:3928}, referred to as the \emph{dynamic power}, and the leakage current model of Skadron~\ea~\cite{Skadron:2004:TMM:980152.980157} :
\begin{equation}
 \Pcpu = (1+ \gamma V)\cdot \xi f V^2, \label{eq:def:cpupower}
\end{equation}
where $\gamma$ is a parameter describing the magnitude of the leakage currents due to capacitor-based circuits, $V$ is the supply voltage and $\xi$ is a parameter defining the power requirements of the microprocessor.
It is known that the leakage currents are temperature-dependent~\cite{2014:devogeleer:samos}.
Henceforth, however, we deem the temperature constant throughout our analysis.

\subsection{Execution Time Model}
\label{section:execution_time_model}

The execution time $\Delta t$ of a code sequence, including slack time $\beta$ and time thieves $\ck$ (the time spend by the operating system), can be modeled as:
\begin{equation}
 \Delta t = \cb \left( \frac{1}{f-\ck}+ \beta \right), \label{eq:def:extime}
\end{equation}
where $\cb$ is the number of clock cycles dedicated to the execution of the user program's statements, $\ck$ the average number of clock cycles per time unit lost due to time thieves, and $\beta$ the average amount of slack time per clock cycle.
By definition $f \geq \ck$ since the system can't steal more clock cycles than what is available.

Time thieves, represented by $\ck$ in Equation~\ref{eq:def:extime},  are clock cycles lost due to low-level operations.
These time thieves have higher priority than $\cb$.
Examples of $\ck$ are pipeline stalls due to branch miss-predictions, misaligned memory
accesses, page faults, operation interventions, interrupt handling, operating system routine tasks, etc.
The slack time represented by $\cb\beta$ is the time the microprocessor cannot continue execution as it is waiting for external data, e.g., in the main memory due to cache misses.
Slack time can be addressed with \acf{OOE}, which would scale down $\beta$ (See Section~\ref{sec:out-of-order-execution}).

\subsection{System's Energy Consumption Model}

Inserting the power model and execution time model from Equation~\ref{eq:def:cpupower} and~\ref{eq:def:extime}, respectively, into the definition of the system's energy consumption in quanta time $i$:
\begin{eqnarray}
\Esi	& = & \Psii \cdot \tx_i \nonumber\\
 	& = & (\Pcpui + \Pdropi + \Pback) \cdot \tx_i \nonumber \\
 	& = & \left((1+ \gamma_i V)\cdot \xi_i f V^2 + \Pdropi +\Pback \right) \nonumber\\
 	&   & \quad \quad \quad \quad \cdot~\cbi\left(\frac{1}{f - \cki} + \beta_i  \right).\label{eq:energy}\label{eq:efcr}\label{eq:Esi}
\end{eqnarray}
Here, $\Psii$ is a monotonic increasing function of $f$, whereas $\Dti$ is a monotonic decreasing function of $f$, given that $\{\Pdropi,\Pback,V,\gamma,\xi_i,\ck,\beta\}\in\mathbb{R}^+$.
Note that $\Pback$ and $\cb$ are scaling factors of $\Esi$ and that this implies that the energy consumed during the execution of a piece of code is linearly dependent on its code complexity and background power demands.
Moreover, this also implies that compiler optimization techniques that target code size optimization will directly also lead to an improved energy profile of the code~\cite{Pallister:2013}.
On the other hand, a microprocessor can also reduce energy consumption by parallelizing code execution, increasing power demands but reducing execution time.
Similar observations between the interaction of energy and power consumption were made by Valluri and John~\cite{Valluri2001}.

At this stage, we only apparently observe an hyperbolic relation between energy and frequency.
We have to take into account the relationship between the voltage and the frequency to find a convex analytical relationship between $E$ and $f$.
Such convexity is of interest as there would exist a microprocessor configuration that minimizes the energy consumption for that particular combination of $\{\Pdropi,\Pback,\gamma,\xi_i,\ck,\beta\}$.

\subsection{Voltage/Frequency Relationship}

The following derivation regarding the energy/frequency relationship is similar to Yuki and Rajopadhye~\cite{yuki2013folklore}; however, different frequency and voltage relationships are used, mainly more contemporary, and the leakage current is scaled more realistically.
Note that $\Pb$ and $\Pdrop$ can be arbitrarily large; their values are inherent to the computer system and independent of the microprocessor.
In the remainder of this work it is also assumed that the temperature of the microprocessor remains constant unless otherwise noted.
In practice it was shown by De\,Vogeleer~\cite{2014:devogeleer:samos} that the microprocessor's power requirements show a strong exponential relation with the temperature.
The non-linear temperature effects complicate the microprocessor's temporal power demands considerably.
The temperature has, however, a small impact on the convex behavior of the \efcr~\cite{thesis:kdv:2015}.
Therefore, we omit the temperature effects on the \efcr\ further on.

\begin{figure}
 \label{fig:dfvs}
 \centering
 	\input{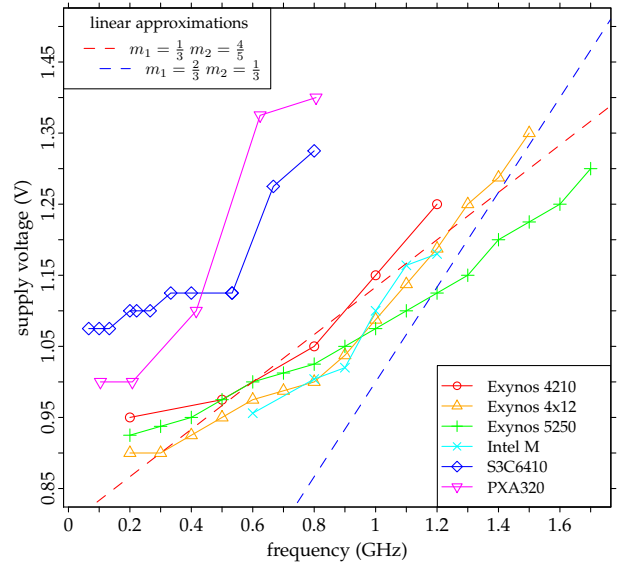}\\\vspace{-1em}
 \caption{Frequency/voltage relationships of multiple modern and vintage application microprocessors, as found in the Linux kernel. Two dashed linear curves are drawn: $V=m_1 f + m_2$; the red is fitted on the depicted data of the three Exynos microprocessors; the blue is borrowed from Yuki and Rajopadhye’s~\cite{yuki2013folklore}.\label{fig:dvfs}}
\end{figure}

For modern microprocessors, the frequency $f$ and supply voltage $V$ are approximately linearly related as shown in Figure~\ref{fig:dvfs}.
It is to be noted that the {\small S3C6410} and the {\small PXA320} are fairly outdated microprocessors and their low performance is visible; the Exynos series and the Intel M are more recent microprocessors designed for embedded multimedia applications, e.g., smartphones and tablets.
The exact relationship between the voltage and frequency is dependent on the physical abilities of the microprocessor's internals, but also on the capability of the microprocessor's voltage and frequency regulator to scale the voltage and frequency on-demand.
When the frequency of a microprocessor is ramped up, the transistors inside need to switch faster to meet timing and delay constraints.
As subparts of transistors are essentially very small capacitors as well; a finite time is required to switch the transistor from one state to another.
Thus if stringent timing delays need to be met, the microprocessor voltage needs be increased accordingly.
The higher voltage supply will decrease the transistors' transition time and capacitors' charging time.
This translates in a positive slope of the frequency/voltage relationship.

An affine transformation between voltage and frequency is expressed as follows:
\begin{equation}
 V = m_1 f + m_2 ,\label{eq:V:f:relationship}
\end{equation}
where $m_1$ and $m_2$ are positive regression coefficients.
Figure~\ref{fig:dvfs} shows the voltage and frequency relationship for several microprocessors.
The values $m_1=\frac{2}{3}$ and $m_2=\frac{1}{3}$, for the dashed blue line in Figure~\ref{fig:dvfs}, are motivated to be adequate for high-performance microprocessors based on theoretical values~\cite{yuki2013folklore}.
Here, the values $m_1=\frac{1}{3}$ and $m_2=\frac{4}{5}$ are shown to better represent the voltage/frequency relationship for microprocessors for embedded applications.
These values are approximates of a linear fit on the combined data of the Exynos microprocessors.

Henceforth, the microprocessor's \emph{default clock frequency window} ($\Fc$) is defined as the clock frequency range bounded by the minimum and maximum clock frequency of the microprocessor:
\begin{equation}
	\Fc = f_\minn \leq \fc \leq f_\maxx.
\end{equation}
We have seen in Section~\ref{section:execution_time_model} that $f \leq \ck$.
Hence, the \emph{exploitable clock frequency window} ($\Fe$) is defined as the frequency range with an upper bound characterized by the microprocessor's maximum frequency $f_\maxx$, and the lower bound defined by the largest of the microprocessor's minimum frequency $f_\minn$ and $\ck$:
\begin{equation}
	\Fe = \max(f_\minn,\ck) \leq \fc \leq f_\maxx.
\end{equation}
It is the exploitable clock frequency window that is open for energy optimization via clock frequency scaling.

\subsection{Optimal Microprocessor Clock Frequency $\fo$}

The power model independent of $V$ is obtained by inserting Equation~\ref{eq:V:f:relationship} in the definition of $\Pcpu$:
\begin{eqnarray}
\Pcpu & = & (1+ \gamma V) \xi f V^2 \nonumber\\
    & = &  a f^4+ b f^3 + c f^2 + d f, \label{eq:Psi}
\end{eqnarray}
where $a = \gamma \xi m_1^3$, $b = m_1^2 \xi (1 + 3\gamma m_2)$, $c = m_1 m_2 \xi ( 3\gamma m_2 + 2 )$, and $d = m_2^2 \xi (\gamma m_2 + 1)$.
This power formulation can then be inserted in the energy consumption model $\Es$ of Equation~\ref{eq:Esi}.

For further analysis the normalized energy consumption $\En$ for code size and background power-independent analysis is introduced.
The normalized energy consumption is defined as
\begin{equation}
 \En = \frac{\Es - \Pback\Delta t}{\cb}.\label{eq:Es:norm}
\end{equation}
Normalizing the energy consumption $\Es$ has no effect whatsoever on its tentative convex properties as $\cb$ and $\Pback$ merely induce an affine transformation of $\En$ without rotation.
$\Pback$ has an effect on the convex properties.
$\Pback$ should however not be part $\En$ as this power component will be present in the system regardless of what the microprocessor is doing.
As a consequence, $\Pback$ should not influence optimal operating settings of the microprocessor.

%

The energy function in Equation~\ref{eq:Es:norm} is called strictly convex over the exploitable clock frequency window if and only if (iff)
\begin{multline}
        \forall f_1 \neq f_2 \in \Fc, \forall t \in (0, 1):\\ \En(tf_1+(1-t)f_2) < t \En(f_1)+(1-t)\En(f_2).
\end{multline}
In other words, if $\Es$ is strictly convex, then $\Es$ possesses no more than one minimum in the exploitable frequency window.
If the minimum of $\Es$ is not within the microprocessor's boundaries, then the minimum $\fo$ can be found via the first derivative of $\En$, while its second derivative must remain positive:
\begin{equation}
 \left(\frac{\partial \En}{\partial f}\right)_{f=\fo} = 0 \quad \text{and} \quad \frac{\partial^2\En}{\partial f^2} > 0.\label{eq:convexity:requirements}
\end{equation}

To simplify the derivative calculation for Equation~\ref{eq:Esi}, $\En$ is split into a polynomial and non-polynomial part, namely $\EnA$ and $\EnB$:
\begin{subequations}
\begin{align}
	\En  & = \EnA + \EnB \nonumber \\
        \EnA & = (a f^4 +b f^3 + c f^2 + d f + \Pdropi ) \cdot \beta \\
        \EnB & = (a f^4 +b f^3 + c f^2 + d f + \Pdropi ) \cdot \frac{1}{f-\ck},
\end{align}\label{eq:Es:split}%
\end{subequations}
The respective derivatives are then as follows:
\begin{subequations}
  \begin{align}
        \frac{\partial \EnA}{\partial f} & = (4a f^3 + 3b f^2 + 2c f + d) \cdot \beta \\
        \frac{\partial \EnB}{\partial f} & = \frac{3af^{4}+\left(2b-4a\ck\right)f^{3}+\left(c-3b\ck\right)f^{2}}{{\left(f-\ck\right)}^{2}}\nonumber\\
        &\quad\quad\quad\quad -\frac{2c\ck f+\Pb+d\ck}{{\left(f-\ck\right)}^{2}} \\
        \frac{\partial^2 \EnA}{\partial f^2} & = (12a f^2 + 6b f + 2c) \cdot \beta \\
        \frac{\partial^2 \EnB}{\partial f^2} & = \frac{6af^{4}+\left(2b-16a\ck\right)f^{3}+\left(12a\ck^{2}-6b\ck\right)f^{2}}{{\left(f-\ck\right)}^{3}} \nonumber\\
        & \quad\quad\quad\quad + \frac{6b\ck^{2}f+2(\Pb+c\ck^{2}+d\ck)}{{\left(f-\ck\right)}^{3}}.
  \end{align}\label{eq:Es:derivatives}%
\end{subequations}%
These equations will be used further on in Section~\ref{sec:sensitivity_of_the_convexity_model} on parameters sensitivity analyses and are also the base for the next section's approximate solutions.

Convex properties can be observed for $\En$.
For $f \rightarrow^+ \ck$, $\EnA$ will approach $\beta\Pdropi$, whereas $\EnB$ is amplified, and tends to positive infinity because of the presence of $f-\ck$ in the denominator.
When $\frac{f}{2} < \ck$, the system is spending more energy in overhead than in the actual program, as the overhead has priority over the program.
In the limit, $\En$ goes to infinity at $\ck$.
At this point the system is overloaded and is not reactive anymore from the point of view of $\cb$.
For $f \rightarrow \infty$, it is $\EnA$ that inflates whereas $\EnB$ approaches zero.
In other words, for the smaller clock frequencies, by virtue of the increased execution time, more energy due to leakage currents needs to be accounted for.
The execution time for large frequencies are dramatically lower, but the dynamic power consumption of the microprocessor increases cubically and the leakage currents increase quartically with clock frequency.
As a result, the convex minimum of the energy function, at the optimal frequency $\fo$, is the point where a balance is found between the consequences of the inflated execution time and the total power demands of the microprocessor.

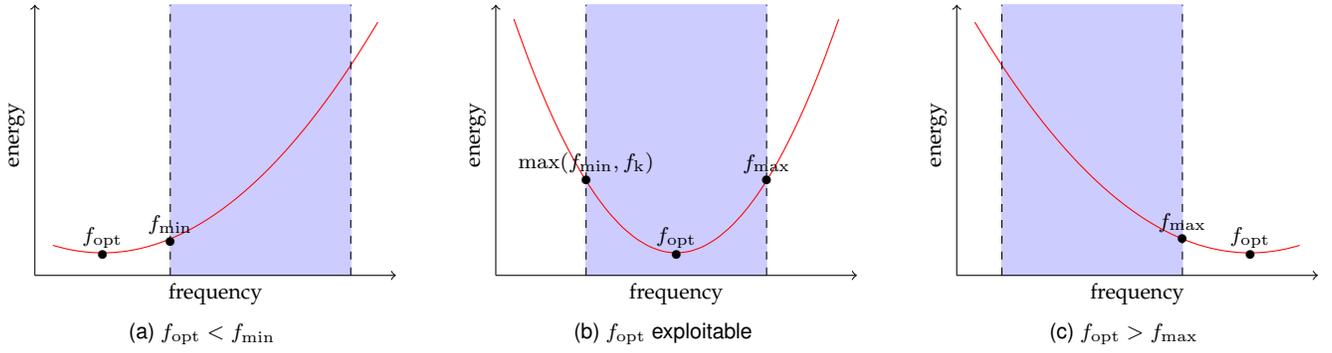
\begin{figure*}
\centering
{\footnotesize
 \subfloat[][$\fo < f_\minn$]{
   \begin{tikzpicture}[scale=1.2]
      
      \filldraw[draw=transparent,color=blue!20] (1.5,0) rectangle (3.5,3);
      
      \draw[scale=0.05,domain=0.2:3.8,smooth,variable=\x,red] plot ({\x/0.05},{5+5.5*(\x-0.75)^2});
      
      \draw[dashed] (1.5,0) -- (1.5,3);
      \draw[dashed] (3.5,0) -- (3.5,3);
      
      \draw (0.75,0.225) node[above] {$\fo$};
      \draw (0.75,0.225) node {\textbullet};
      
      \draw (1.5,0.375) node[above] {$f_\minn$};
      \draw (1.5,0.375) node {\textbullet};
      
      \draw[->] (0,0) -- (4,0);
      \draw[->] (0,0) -- (0,3);
      \draw (2,0) node[below] {frequency};
      \draw (-0.2,2) node[left,rotate=90] {energy};
      
\end{tikzpicture}
 }\hspace{2em}
 \subfloat[][$\fo$ exploitable]{
   \begin{tikzpicture}[scale=1.2]
      
      \filldraw[draw=transparent,color=blue!20] (1,0) rectangle (3,3);
      
      \draw[scale=0.05,domain=0.2:3.8,smooth,variable=\x,red] plot ({\x/0.05},{5+16*(\x-2)^2});
      
      \draw (2,0.225) node[above] {$\fo$};
      \draw (2,0.225) node {\textbullet};
      
      \draw (1,1.05) node[above] {$\max(f_\minn,\ck)$};
      \draw (1,1.05) node {\textbullet};
      
      \draw (3,1.05) node[above] {$f_\maxx$};
      \draw (3,1.05) node {\textbullet};
      
      \draw[dashed] (1,0) -- (1,3);
      \draw[dashed] (3,0) -- (3,3);
      
      \draw[->] (0,0) -- (4,0);
      \draw[->] (0,0) -- (0,3);
      \draw (2,0) node[below] {frequency};
      \draw (-0.2,2) node[left,rotate=90] {energy};
      
\end{tikzpicture}
 }\hspace{2em}
 \subfloat[][$\fo > f_\maxx$]{
   \begin{tikzpicture}[scale=1.2]
      
      \filldraw[draw=transparent,color=blue!20] (0.5,0) rectangle (2.5,3);
      
      \draw[scale=0.05,domain=0.2:3.8,smooth,variable=\x,red] plot ({\x/0.05},{5+5.5*(\x-3.25)^2});
      
      \draw[dashed] (0.5,0) -- (0.5,3);
      \draw[dashed] (2.5,0) -- (2.5,3);

      \draw (3.25,0.225) node[above] {$\fo$};
      \draw (3.25,0.225) node {\textbullet};
      
      \draw (2.5,0.4) node[above] {$f_\maxx$};
      \draw (2.5,0.4) node {\textbullet};
      
      \draw[->] (0,0) -- (4,0);
      \draw[->] (0,0) -- (0,3);
      \draw (2,0) node[below] {frequency};
      \draw (-0.2,2) node[left,rotate=90] {energy};
      
\end{tikzpicture}
 }}
 \caption{The location of the optimal frequency $\fo$ w.r.t. default clock frequency window (blue) is an indication of which energy optimization technique is most effective: (a) when $\fo$ is left of the exploitable clock frequency window ($\fo < f_\minn$), one should set the clock frequency as low as possible; (b) if $\max(f_\minn,\ck) < \fo < f_\maxx$ then chasing $\fo$ will yield the best energy efficiency; (c) when $\fo > f_\maxx$, then the race-to-halt energy optimization technique is most effective. Powerful microprocessors are most likely to fall in the category (c), e.g. DGEMM 8C in Figure~\ref{fig:Hager}, whereas low-power microcomputers are more likely to be in category (b), e.g. TI C62 in Figure~\ref{fig:Senn}. \label{fig:convexity:types}}
\end{figure*}

Given an energy/frequency convex behavior, three classes of microprocessor configurations can be distinguished, as shown in Figure~\ref{fig:convexity:types}.
When the optimal clock frequency $\fo$ is left of the default clock frequency window ($\fo < f_\minn$), setting the clock frequency at $f_\minn$ yields the best energy gains; if $\max(f_\minn,\ck) < \fo < f_\maxx$ then chasing $\fo$ will earn the best energy efficiency; and when $\fo > f_\maxx$, then the race-to-halt\footnote{The energy optimization technique \emph{race-to-halt} runs the microprocessor at full speed until all tasks are completed; then the microprocessor is put in a low-power mode.} energy optimization technique is shown to be most effective.
It was noted by Rizvandi~\cite{5493460} that under certain circumstances it can be more efficient, in terms of energy consumption, to have a binary frequency scheme, including the maximum and minimum clock frequency, rather than scaling the clock frequency through the whole frequency space.
The presented performance-oriented work, and also the user-oriented work of Seeker~\ea~\cite{DBLP:conf/iiswc/SeekerPLF14}, suggest that this is, in fact, not the case.
$\fo$ may assume any frequency within the default clock frequency window, and may fluctuate throughout the code execution depending on the kind of operations scheduled.

\subsection{Approximate Optimal Clock Frequency $\fo$}

The power model (Equation~\ref{eq:def:cpupower}) in the energy consumption formulation of Equation~\ref{eq:Esi} is of the fourth order.
When the fourth-order power equation can be adequately approximated with a quadratic polynomial, the derivations can be simplified somewhat.
The power consumption $\Ps$ of the system can then be represented as:
\begin{equation}
  \Ps = a f^4+ b f^3 + c f^2 + d f \approx k f^2 + l f + m, \label{eq:Ps:approx}
\end{equation}
and accordingly the energy consumption of the system becomes
\begin{equation}
  \En = ( k f^2 + l f + m   + \Pdropi ) \cdot \Delta t. \label{eq:Es:approx}
\end{equation}
$k\in\mathbb{R}^+_0$, though $\{l,m\}\in\mathbb{R}$.
The first and second derivatives of the normalized energy consumption are then as follows:
\begin{subequations}
  \begin{align}
        \frac{\partial \En}{\partial f} & = \beta (2 k f + l) - \frac{\ck(2 k f + l) -k f ^2 + m + \Pdropi}{(f-\ck)^2}, \\
        \frac{\partial^2 \En}{\partial f^2} & = 2k \beta + \frac{2(\ck^2 k + \ck l + m + \Pdropi)}{(f-\ck)^3} .
        \end{align}\label{eq:Es:derivatives:simplified}%
\end{subequations}
There exists a convex minimum if $\frac{\partial \En}{\partial f}$ has a root and $\frac{\partial^2 \En}{\partial f^2}$ is a monotonous increasing function.
In other words:
\begin{eqnarray}
        0  & = & 2 k \beta f^3   + (k+\beta (l-4 \ck k)) f^2 + 2 \ck \beta (\ck k -l)f \nonumber\\
          &   & \hspace{0em} - 2 \ck k f - ( m + \Pdropi + \ck l(1 - \beta\ck)) \label{eq:Es:approx:roots}\\
         2k \beta & \geq & -2\frac{\ck^2 k + \ck l + m + \Pdropi}{(f-\ck)^3}. \nonumber
\end{eqnarray}
The solution to Equation~\ref{eq:Es:approx:roots} is the frequency that minimizes energy consumption.
Via Ferarri's solution~\cite{11820sdsdsd65} for the calculation of the roots of a third order polynomial, the optimal frequency can be determined analytically.
Yet, the analytical formulation to calculate the roots of a cubic polynomial is still elaborate.
Let's assume some further simplifications.
For $\beta = 0$, one gets that
\begin{eqnarray}
  \fo & = & \ck + \frac{\sqrt{2k^2  \ck^2 + 2 k ( m + \Pdropi + \ck l)}}{k} \\
  0 & \leq & 2\frac{\ck^2 k + \ck l + m + \Pdropi}{(f-\ck)^3} \nonumber.
\end{eqnarray}
If all parameters are elements of $\mathbb{R}^+$, the latter inequality holds whenever $\ck < f$.
Additionally, for $\ck = 0$, one obtains
\begin{eqnarray}
        \fo & = & \sqrt{\frac{m + \Pdropi}{k}} \\
          0 & \leq & \frac{2(m + \Pdropi)}{f^3}, \nonumber
\end{eqnarray}
which is only valid for $-\Pdropi < m$.
These simplified models for $\beta = \ck = 0$ may be used when the context allows for, i.e., when $\cb$ is executed without any interruption.
For example, from practical experience and in the literature, $\ck$ is often observed to be close to zero in a multi-core context.
$\beta$ may vary considerably for different applications and should be assessed before deeming insignificant.

\section{Experimental Results}
\label{chapter:2b:sec:experimental:results}

In this section experimentally-obtained power and execution time measurement traces are presented and used as a reference to study the \efcr\ in the next section.

\subsection{Platform and Benchmark Description}

A Samsung Galaxy S2, sporting an ARM Cortex A9 dual-core microprocessor, was used as testbed.
The A9 uses clock frequency ranges from 0.2\,GHz to 1.6\,GHz in steps of 100\,MHz.
The Gold-Rader implementation of the bit-reverse algorithm was used as benchmark; it is part of the ubiquitous \ac{FFT} algorithm, in which it rearranges deterministically elements in an array.
Besides the Gold-Rader algorithm, the \beebs\ benchmark~\cite{DBLP:journals/corr/PallisterHB13} was also run on an \odroidxue, featuring an Exynos 5240, while the execution time and power was measured.
The measurement data of the Gold-Rader algorithm and \beebs\ show large similarities.
The Gold-Rader algorithm is chosen as a base for the expound in the sequel.
More info on the \beebs\ measurements can be found in De\,Vogeleer~\cite{thesis:kdv:2015}.

%

\subsection{Execution Time and Power Consumption}

\begin{figure*}[t]
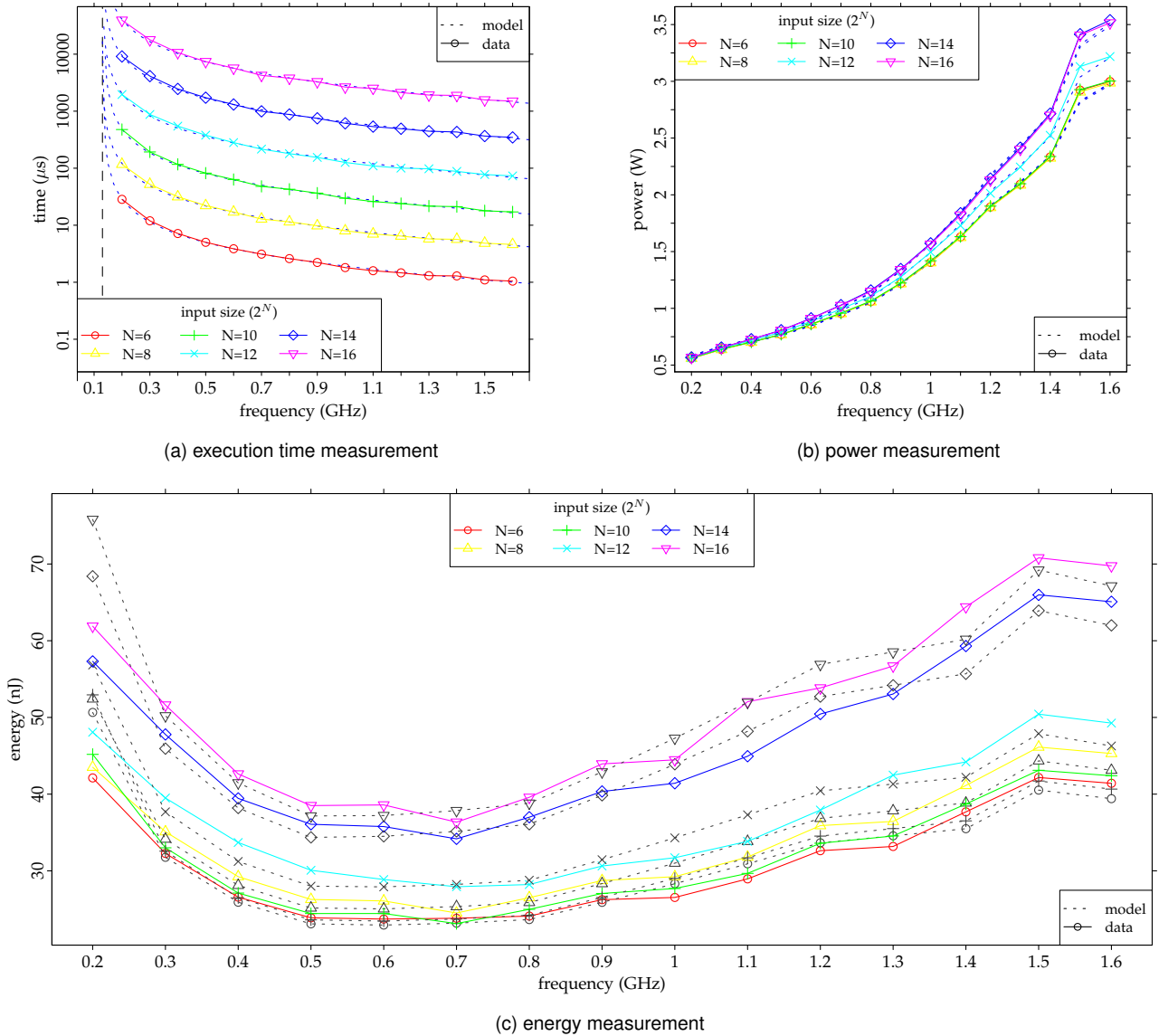

    \centering
	\subfloat[][execution time measurement]{\input{measurement-time-A9.tex}\label{fig:benchmark:measurements:time}}\hspace{1em}
	\subfloat[][power measurement]{\input{measurement-power-A9.tex}\label{fig:benchmark:measurements:power}}\\
	\subfloat[][energy measurement]{\input{measurement-energy-A9.tex}\label{fig:benchmark:measurements:energy}}
	\caption{Experimental data for the Cortex A9 microprocessor. The energy consumption of the benchmarks with different input sizes is shown for the Gold-Rader algorithm. The solid lines represent the measured data whereas the dotted lines is the product of the fitted power and execution time models from Figure~\ref{fig:benchmark:measurements:time} and Figure~\ref{fig:benchmark:measurements:power}, respectively.}
\end{figure*}

Figure~\ref{fig:benchmark:measurements:time} shows the execution time of the Gold-Rader algorithm on the A9 microprocessor.
Table~\ref{table:measurement:coefficients:time} shows the fitted execution time parameters as per Equation~\ref{eq:def:extime}.
 \begin{table}
    \caption{Benchmark execution time model parameters: $\xi$, $\gamma$ and $\Ps$ as per Equation~\ref{eq:def:cpupower}, and $\cb$, $\ck$, $\beta$ as per Equation~\ref{eq:def:extime} for running the Gold-Rader algorithm on the A9 microprocessor.
    These values were used for the fitted models in Figure~\ref{fig:benchmark:measurements:power}.\label{table:measurement:coefficients:time}}
   \begin{center}
     \begin{tabular}{c|c|c|c|c|c|c|}\cline{2-7}
     & \multicolumn{6}{c|}{ {\sc Gold-Rader -- input size} 2$^N$  -- A9 } \\\hline
  \multicolumn{1}{|c|}{$N$}  &  6  &  8  &  10  &  12  &  14  &  16  \\\hline
  \multicolumn{1}{|c|}{$\cb$} &  1.943  &  8.596  &  31.1  &  144.359  &  670.8  &  2918.837  \\
  \multicolumn{1}{|c|}{$\ck$} &  0.134  &  0.129  &  0.137  &  0.13  &  0.13  &  0.129   \\
  \multicolumn{1}{|c|}{$\beta$} &  -0.166  &  -0.167  &  -0.152  &  -0.202  &  -0.183  &  -0.182   \\\hline
 \multicolumn{1}{|c|}{$\xi$} &  0.101  &  0.108  &  0.134  &  0.137  &  0.44  &  0.011  \\
 \multicolumn{1}{|c|}{$\gamma$} &  5.578  &  5.127  &  4.030  &  4.36  &  1.035  &  65.985   \\
 \multicolumn{1}{|c|}{$\Ps$} &  0.480  &  0.480  &  0.477  &  0.469  &  0.394  &  0.407   \\
\hline 
     \end{tabular}
    \end{center}
 \end{table}
The fitted execution time model has a relative error such that 90\,\% of the errors are between 0.18\,\% and 7.36\,\% and shows a median of 3.12\,\% for the execution time traces.


Figure~\ref{fig:benchmark:measurements:power} shows the power profile of the Gold-Rader algorithm on the A9.
All traces were recorded while the temperature of the hardware fluctuated.
During the recording of the power traces the temperature of the testbed was artificially oscillated around 37$^\circ$C and then the power samples at a temperature of 37$^\circ$C were selected.

Table~\ref{table:measurement:coefficients:time} also shows the fitted values for $\xi$, $\gamma$ and $\Ps$ as per Equation~\ref{eq:def:cpupower} for the A9.
Discrete voltage/frequency pairs were used to fit the measured data as reported in Figure~\ref{fig:dvfs} for the Exynos 4210.
%

The fitted model parameters in Table~\ref{table:measurement:coefficients:time} seem to be consistent for an input size up to $2^{12}$.
The fitted model parameters for larger input sizes seem to be much different.
Note that array sizes up to $2^{9}$ fit in the L1 cache, while sizes over $2^{18}$ are too big to fit in the L2 cache.
Therefore external memory accesses and microprocessor slack time may influence the power of the microprocessor.
Overall, the power variation of the different input sizes are not as large as what was observed for the case of the execution time.
The magnitude of the power of all traces are all of the same order, whereas for the execution time it may differ by multiple orders.

As observed from Figure~\ref{fig:benchmark:measurements:power} the power model fits well on the experimental data.
The fitting errors for the A9 are between 0.07\,\% and 3.18\,\% with a median of 0.86\,\%.
The fitted model for the A9 in Figure~\ref{fig:benchmark:measurements:power} for $f = 1.5$\,GHz seems to deviate persistently from the measured data.
This could be due to a slightly higher supply voltage at 1.5\,GHz than reported in Figure~\ref{fig:dvfs} for the Exynos 4210 microprocessor.

\subsection{Energy Consumption}
\label{sec:energy_consumption}

The estimated experimental energy consumptions are obtained by multiplying the power traces with the execution time traces for each frequency.
This was done for both the experimental traces and the fitted power and execution time models.
Figure~\ref{fig:benchmark:measurements:energy} shows the energy consumption of the Gold-Rader algorithm on the A9 microprocessor.
The fitted errors are the sum of the errors of the power and execution time traces separately.
For the A9 traces a clear minimum energy consumption is observed between 500\,MHz and 800\,MHz.

\section{Sensitivity of the Convexity Model}
\label{sec:sensitivity_of_the_convexity_model}

To analyze the behavior and parameter sensitivity of the convexity model of Equation~\ref{eq:efcr}, the Cortex A9 processor of the Exynos 4210 is used as reference use case, representative for embedded multimedia applications, e.g., smartphones~\cite{paper:karel:warsaw}.
The following values were used, based on the measurements presented in the previous section: $m_1$ = 0.330 [V/f],  $m_2$ = 0.808 [V], $\beta$ = 0 [s], $\gamma $ = 3.137 [V$^{-1}$], $\ck$ = 0.130 [GHz], $\xi_\maxx$ = 0.181 [W/(GHz$\cdot$V$^2$)], $\xi_\minn$= 0.155 [W/(GHz$\cdot$V$^2$)], $\Pdrop$ = 0 [W]. The microprocessor's clock frequency starts at 200\,MHz and goes to 1.6\,GHz and $\xi$ is a parameter that describes the power profile of an application.
The values for $\beta$, $\ck$, $\gamma$ and $\xi$ were defined via fitting as presented in the previous sections.
The microprocessor's clock frequency is also considered a continuous variable from here on.
In reality the clock frequency is limited to a discrete set of values.
However, for analytical purposes, not to mention the aesthetics of the graphs, the clock frequency is deemed continuous.

In the next sections we will look at how time thieves and \ac{OOE} impacts the convexity model.
Time thieves are basically clock cycles lost to overhead, whereas \ac{OOE} is an intelligent instruction execution scheme to minimize execution slack time.

\subsection{What About Those Time Thieves?}

When considering the execution time of a code sequence, $\ck$ was previously defined as the number of clock cycles per time unit not available to the execution of the user code.
These clock cycles are spent, for example, to handle microprocessor exceptions, or to execute operating system routine tasks.
$\ck$ can therefore be regarded as little \emph{time thieves}.
From a mathematical point of view, the presence of $\ck$ in Equation~\ref{eq:efcr} also introduces some complexity for derivations such as Equation~\ref{eq:Es:derivatives}.
Bear in mind that the microprocessor's clock frequency $f$ is always larger than $\ck$; otherwise the execution time is not defined.
Consequently, $\ck < f_\maxx$ must be satisfied.

\begin{figure}
  \begin{center}
	\subfloat[][$\fo$($\ck$,$\xi$)]{
	  \input{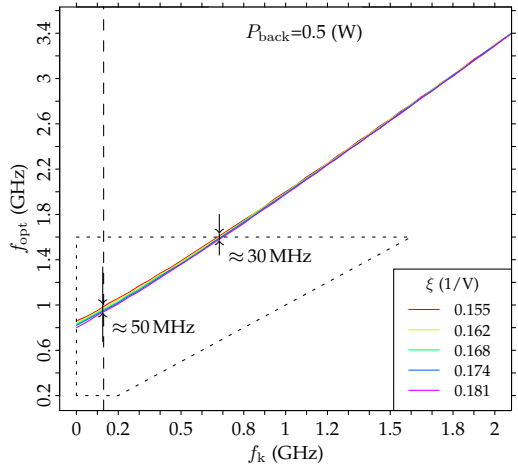}
	  \begin{tikzpicture}[overlay]
	    \draw[->] (-4.80,3.50) -- (-4.80,3.24);
	    \draw     (-4.80,3.24) -- (-4.80,3.15);
	    \draw[<-] (-4.80,3.15) -- (-4.80,2.95);
	    \draw (-4.80,2.95) node[right] {{\scriptsize $\approx$\,30\,MHz}};
	    
	    \draw[->] (-6.350,1.80) -- (-6.350,2.19);
	    \draw     (-6.350,2.19) -- (-6.350,2.30);
	    \draw[<-] (-6.350,2.30) -- (-6.350,2.80);
	    \draw (-6.350,2.00) node[right] {{\scriptsize $\approx$\,50\,MHz}};
	  \end{tikzpicture}
	 \label{fig:4a}}\\
	\subfloat[][$\fo$($\ck$,$\Pb$)]{\input{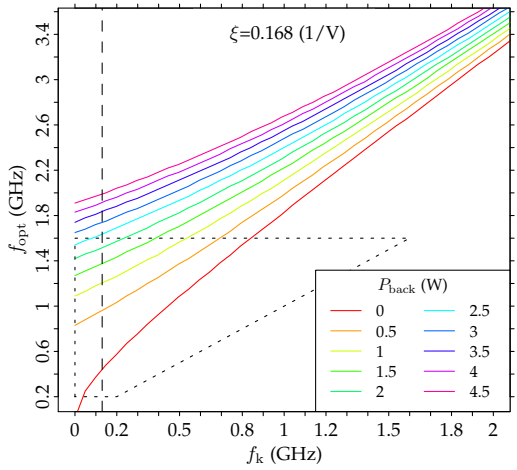}}
	\caption[Optimal frequency for variable level of $\ck$ in function of $\xi$ and $\Pb$]{Optimal microprocessor frequency $\fo$ for variable levels of $\ck$ in function of $\xi$, on the top, and $\Pb$, on the bottom. A typical value for $\ck$ is drawn at 0.13\,GHz (dashed vertical line). The area encapsulated by the dotted line signals the microprocessor's exploitable clock frequency window: $\max(0.2\,\text{GHz},  \ck) \leq f \leq 1.6\,\text{GHz}$.\label{fig:sensitivity:cck}}
  \end{center}
\end{figure}

Figure~\ref{fig:sensitivity:cck} shows the sensitivity of $\ck$ with regards to the optimal frequency $\fo$, the microprocessor power ($\Pcpu \propto\xi$), and the background power $\Pb$.
In the bottom plot it is seen that $\fo(\ck=0,\Pb=0.5)\approx 0.8$\,GHz.
The optimal frequency increases for increasing values of $\ck$ and hits the microprocessor's maximum frequency $f_\maxx=1.6$\,GHz around $\ck=0.7$\,GHz.
At this point, about 45\,\% ($\approx$\,0.7/1.6) of the clock cycles would not be available to the code sequence.
Furthermore, it is observed that $\fo > \ck$ always holds.
The effect of the microprocessor's power demands on $\fo$ is fairly small, expressed by the $\xi$ parameter.
A 30\,MHz to 50\,MHz difference in $\fo$ is observed between the minimum and maximum microprocessor's power usage as $\xi$ varies between 0.155\,V$^{-1}$ and 0.181\,V$^{-1}$ (see Figure~\ref{fig:4a}).

The background power usage $\Pb$ has a bigger impact on $\fo$ than $\xi$.
For $\Pb = 0$,  $\fo$ even drops below the minimum operation frequency of the microprocessor.
Increasing $\Pb$ inflates $\fo$.
For $\ck=0$ and $\Pb\approx 2.5$\,W the optimal frequency already surpasses $f_\maxx$.
For a typical value of $\ck$ (130\,MHz), an increase in $\fo$ is observed for increasing values of $\Pb$; yet, the increase becomes smaller for larger values of $\Pb$.
The average difference between $\fo(\ck=0)$ and $\fo(\ck=0.13)$, within the microprocessor's clock frequency range, is approximately 100\,MHz.

In the rest of this section it will be assumed for simplicity that $\ck \ll f$ unless otherwise stated.
For a more realistic estimate of $\fo$, in case $\ck$ is not negligible, it was observed from the graphs that adding 100\,MHz to $\fo$ is a reasonable assumption.

\subsection{Absence of Time Thieves}

It is not unthinkable that, in particular contexts, $\ck$ is indeed negligibly small compared to $f$: $\ck \ll f$.
For example, such occasions may occur when the clock frequency microprocessor is reasonably fast, or the code sequence of concern is running only on one of the available cores of a multi-core microprocessor without interruption.
Assuming $\ck$ negligible considerably simplifies Equation~\ref{eq:Es:derivatives}.
For $\max(f_\minn,\ck) < \fo < f_\maxx$, $\En$ was said to be strictly convex iff there exists only one point in the exploitable clock frequency window for which $\frac{\partial \En}{\partial f}=0$ and $\frac{\partial^2 \En}{\partial f^2} > 0$.
Given the system of Equations~\ref{eq:Es:derivatives}, these two requirements translate, respectively, into:
\begin{multline}
    \frac{\Pb}{\fo^2} = 4a\beta \fo^3 + 3 (a + b\beta ) \fo^2 \\+ 2 ( b + c \beta ) \fo + (d\beta + c ), \label{eq:time:thieves:requirement:zero}
\end{multline}
\begin{multline}
    0 < 12a\beta \fo^2 + 6(a+ b\beta ) \fo \\ + 2 (b + c\beta ) + 2\frac{\Pb}{\fo^3}.\label{eq:time:thieves:requirement:possitive}
\end{multline}
Recall that for all constants in this system of equations: $\{a,b,c,d,\beta\} \in \mathbb{R}^+$.
Thus the requirement in Equation~\ref{eq:time:thieves:requirement:possitive} is satisfied by default as the right-hand side will never be negative.
Accordingly, the root requirement of Equation~\ref{eq:time:thieves:requirement:zero} is also satisfiable.
It is immediately clear that the background power demands $\Pb$ directly controls the optimal frequency $\fo$.
The constants $\{a,b,c,d\}$ describe the microprocessor's power usage whereas $\Pb$ describes the power demands of everything in the computer system besides the microprocessor.
For systems with a large $\Pb$, e.g., servers or desktop computers, $\fo$ will therefore be higher than for systems with a low $\Pb$, e.g., wireless sensors.
Moreover, $\fo$ may be so high that it is larger than the maximum microprocessor's clock frequency.

Figure~\ref{fig:sensitivity:Pback} shows the optimal frequency for a variable background power consumption $\Pb$ and microprocessor loads $\xi$.
\begin{figure}
  \begin{center}
	\subfloat[][$\fo(\Pback,\xi)$]{
	  \input{sensitivity-Pback-fopt}
	  \begin{tikzpicture}[overlay]
	    \draw[->] (-4.16,4.25) -- (-4.16,4.5);
	    \draw     (-4.16,4.5) -- (-4.16,4.75);
	    \draw[<-] (-4.16,4.75) -- (-4.16,5);
	    \draw (-4.16,4.4) node[right] {{\scriptsize $\approx$\,100\,MHz}};
	    
	    \draw[->] (-4.955+0.075,4.62) -- (-4.465+0.075,4.621);
	    
	    \draw     (-4.465+0.075,4.62) -- (-3.975+0.075,4.62);
	    
	    \draw[<-] (-3.97+0.075,4.62) -- (-3.550+0.075,4.62);
	    \draw (-5.1+0.075,4.61) node[above] {{\scriptsize $\approx$\,0.5\,W}};
	  \end{tikzpicture}
	}\\
	\subfloat[][$\Pcpu$/$\Pb$ ratio at $\fo$]{
	  \input{sensitivity-Pback-Pcpu}
	  \begin{tikzpicture}[overlay]
	    \draw[->] (-4.235+0.075,4.7) -- (-4.235+0.075,4.95);
	    \draw     (-4.235+0.075,4.95) -- (-4.235+0.075,5.1);
	    \draw[<-] (-4.235+0.075,5.1) -- (-4.235+0.075,5.35);
	    \draw (-4.235+0.075,5.35) node[left] {{\scriptsize $\approx$\,0.05}};
	  \end{tikzpicture}
	}
	\caption[Optimal frequency for variable background power consumption $\Pb$]{Optimal microprocessor frequency $\fo$ for variable background power consumption $\Pb$. On the top $\fo$, is shown for various microprocessor loads $\xi$. On the bottom, the ratio between the background power and the microprocessor power $\Pcpu$ at $\fo$ is shown. The area between the dotted lines signals the effective clock frequency window: $0.2\,\text{GHz} \leq f \leq 1.6\,\text{GHz}$.\label{fig:sensitivity:Pback}}
  \end{center}
\end{figure}
Also, the ratio between the microprocessor $\Pcpu$ and the background $\Pb$ power consumption is given.
The area encapsulated by the dotted line signals the operating range of the microprocessor.
For the microprocessor to be able to exploit the minimum-energy operation frequency, the background power consumption needs to be between 0.02\,W and about 2.75\,W, depending on the exact microprocessor load.
The influence of the different microprocessor loads on $\Pb$ is not significant; at 1.6\,GHz there is a 0.5\,W difference between $\Pb$ for $\xi_\minn$ and $\xi_\maxx$.
If $\Pb$ is larger than 2.75\,W, it is advised to run the microprocessor at the maximum clock frequency to minimize energy consumption.
Under such conditions, the energy optimization technique known as \emph{race-to-halt} is a good strategy.
This was also Yuki and Rajopadhye's~\cite{yuki2013folklore} main conclusion while studying high-performance computers.
The optimal frequency $\fo$ surpasses the microprocessor's maximum frequency roughly around the point where the background power demands become larger than the microprocessor's power usage.
Battery-powered electronic systems such as embedded systems, wireless sensors or smartphones aim at minimizing their background power demands, which thus increases the feasibility of $\fo$ exploitation.
For more powerful computers, however, such as servers, the optimal frequency will be very likely out of reach of the microprocessor's capabilities: $\fo > f_\maxx$.
For example, Seo~\ea~\cite{Soe2012effect} claim that \ac{DVFS} in general hardly improves the energy efficiency of mobile multimedia electronics.
The testbed power measurements of their embedded system show, however, that their $\Pcpu$ to $\Pback$ ratio is smaller than 1 to 18, and their $m_1$ is very small.
For their specific testbed, $\fo$ is very likely  larger than $f_\maxx$, and race-to-halt should indeed be most benificial when aiming for energy savings.

\subsection{Out-of-Order Execution}
\label{sec:out-of-order-execution}

\begin{figure}
  \begin{center}
	\subfloat[][$\fo$($\beta$,$\xi$)]{
	  \input{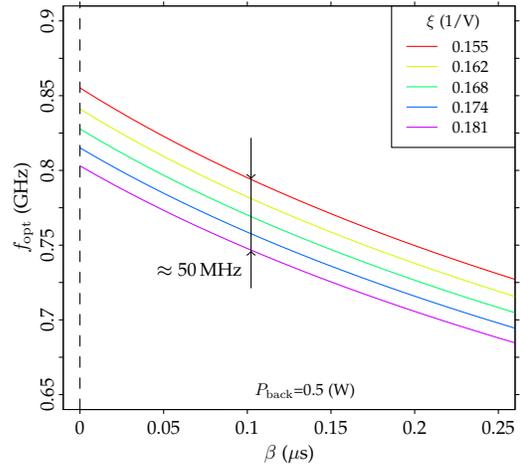}
	  \begin{tikzpicture}[overlay]
	    \draw[->] (-4.425,2.5) -- (-4.425,3);
	    \draw     (-4.425,3) -- (-4.425,3.95);
	    \draw[<-] (-4.425,3.95) -- (-4.425,4.5);
	    \draw (-4.425,2.75) node[left] {{\scriptsize $\approx$ 50\,MHz}};
	  \end{tikzpicture}
	}\\
	\subfloat[][$\fo$($\beta$,$\Pb$)]{\input{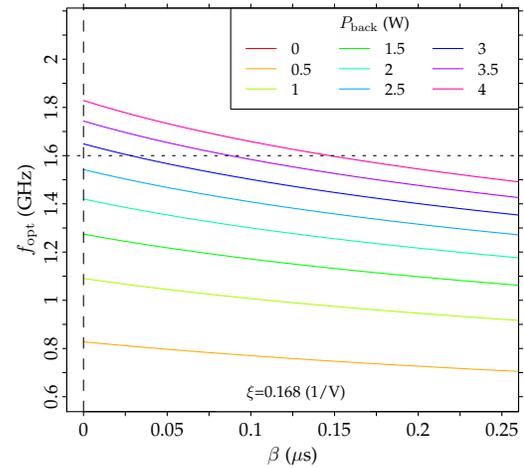}}
	\caption[Optimal frequency for variable levels of $\beta$ in function of $\xi$ and $\Pb$]{Optimal microprocessor frequency $\fo$ for variable levels of $\beta$ in function of $\xi$, on the top, and $\Pb$, on the bottom. The area below the horizontal dotted line signals the microprocessor's default clock frequency window ($0.2$\,GHz $\leq f \leq 1.6$\,GHz).\label{fig:sensitivity:OOE}}
  \end{center}
\end{figure}

Out-of-order execution ({\small OOE}) is parametrized via $\beta \in [0,\infty[$ in Equation~\ref{eq:def:extime}: 
$\beta = 0$ when \ac{OOE} is perfectly able to cover the time during external memory accesses with data-independent code execution; otherwise $\beta$ is larger than 0.
The system's normalized energy consumption, assuming $\ck \approx 0$, is given by:
\begin{equation*}
 \En = ( a f^4+ b f^3 + c f^2 + d f   + \Pback ) \cdot  \left( \frac{1}{f}+\beta\right).
\end{equation*}
Its requirements for convexity are defined the same as for the case where time thieves are absent, given by Equation~\ref{eq:time:thieves:requirement:zero} and~\ref{eq:time:thieves:requirement:possitive}.
It can be observed that for $\beta = 0$ the most left-hand term in Equation~\ref{eq:time:thieves:requirement:zero} becomes zero, resulting in an increased $\fo$ for the equality to be satisfied.
Similarly, the larger $\beta$, the more $\fo$ needs to decrease for the inequality of Equation~\ref{eq:time:thieves:requirement:possitive} to hold.
Figure~\ref{fig:sensitivity:OOE} shows the sensitivity of the $\beta$ parameter on the optimal frequency $\fo$.
Indeed, from the figure, it is observed that $\fo$ decreases for increasing $\beta$.
Moreover, $\fo$ changes about 100\,MHz over a 0 to 0.25\,$\mu$s $\beta$ range for medium levels of $\Pb$.
The larger $\Pb$, the larger the spread in $\fo$ for variable $\beta$.
For $\Pb$ over 4\,W, the $\fo$ spread between $\beta = 0$ and $\beta = 0.25$ increases to more than 200\,MHz.

In theory, $\beta$ can be frequency-dependent as well.
That is, the memory clock frequency can be scaled along with the microprocessor's frequency, this to ensure the timely delivery of data in the microprocessor registries and caches.
$\beta$ in such a case would not be constant over $f$.
Here, it was assumed that the microprocessor's clock frequency, once set at $\fo$, doesn't change over time.
Another common approach to save energy is to have a variable clock frequency to minimize \ac{OOE} slack-time and also energy consumption.

%
%
%

\section{State of the Art}
\label{sec:state_of_the_art}

In the previous sections, it is shown that the energy consumption of a microprocessor shows convex properties with regard to its clock frequency.
The convex energy consumption curve has been mentioned before several times in the literature.
A sensitivity study of the convexity model, as presented here, has not been reported before.
A series of papers, approaching the problem from a chip point of view, without the consideration of software, have shown the energy consumption with respect to \acf{DVFS}~\cite{Snowdon_RH_05,Fan:2003:SPM:2157911.2157927,LeSueur:2010:DVF:1924920.1924921,Seok:2011:PSI:2024724.2024943}.
The literature puts forward some motivation for the energy consumption's convexity, but rarely provides analytical frameworks based on physical explanations.
For example, Senn~\ea~\cite{DBLP:journals/ejasp/SennLJM05} and Austin and Wright~\cite{Austin:2014:MIM:2689710.2689716} provide a heuristic model. 
Other studies, e.g., Hager~\ea~\cite{DBLP:journals/corr/abs-1208-2908} and Freeh ~\ea~\cite{Freeh:2007:AET:1263127.1263246}, discuss what the consequences are of said behavior and how to exploit them, from a high-level point of view.
Other researchers have also shown energy measurements under \ac{DVFS} processes but no convexity is shown by the measurements, e.g., Sinha and Chandrakasan~\cite{Sinha:2001:JWB:378239.378467}, and \v{S}imuni\'c \ea~\cite{782199}, who are not running their benchmarks on top of an \ac{OS}.
Authors, such as Austin and Wright~\cite{Austin:2014:MIM:2689710.2689716} and Snowdon~\cite{Snowdon_RH_05,Fan:2003:SPM:2157911.2157927}, have shown more specifically that 
for applications with certain behavioral patterns no energy convexity is observed.
However, the energy consumption model presented in our work can explain such behavior.

In the \ac{VLSI} design domain, voltage scaling has also been discussed but usually for a fixed frequency~\cite{Zhai:2004:TPL:996566.996798,4271869,1167510}.
The aim of the voltage scaling is to find a minimum energy operation point where the digital circuit yields the correct output.
The major trade-off is between increased circuit latency and leakage power, and decreasing dynamic power.
This trade-off also yields a convex energy consumption curve, but for a fixed frequency.
In this paper, however, the combined effect of voltage/frequency scaling is of interest.

\begin{figure*}
    \begin{center}
	\subfloat[][Fan~\ea \cite{Fan:2003:SPM:2157911.2157927}]{\includegraphics{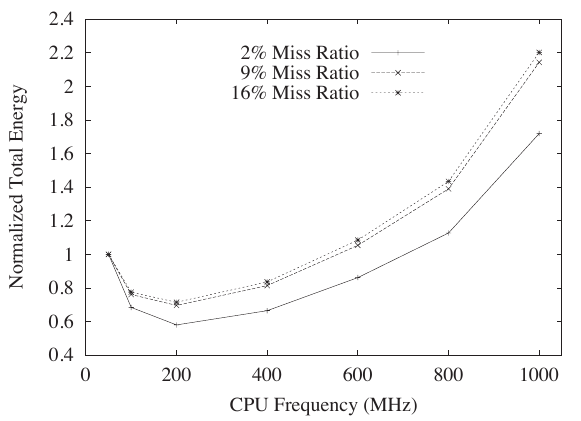}}
	\subfloat[][Le Sueur and Heiser \cite{LeSueur:2010:DVF:1924920.1924921}]{\includegraphics[scale=0.85]{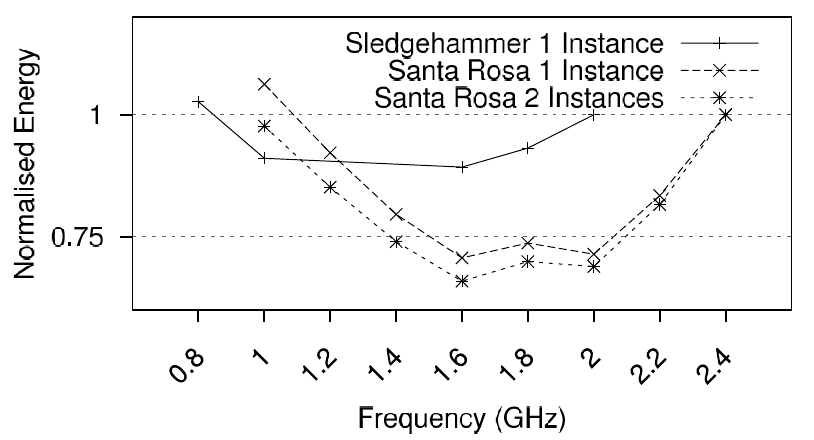}}\\
	\subfloat[][Hager \ea~\cite{DBLP:journals/corr/abs-1208-2908}]{\includegraphics{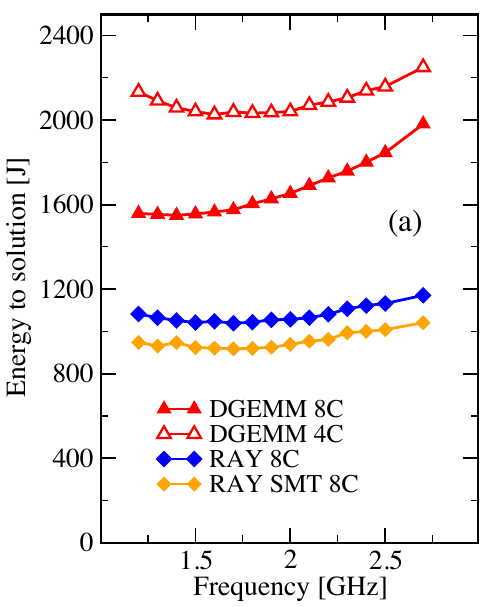}\label{fig:Hager}}
	\subfloat[][Snowdon \ea~\cite{Snowdon_RH_05}]{\includegraphics{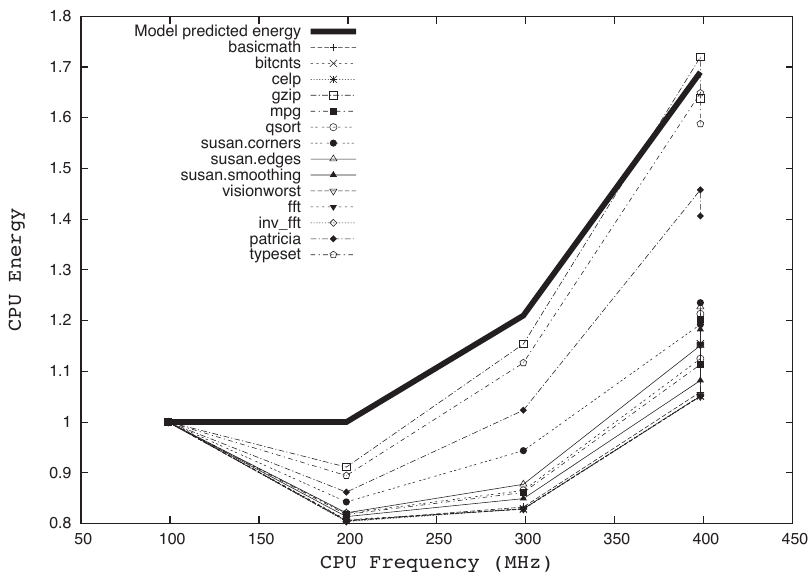}}\\
	\subfloat[][Austin and Wright \cite{Austin:2014:MIM:2689710.2689716}]{\includegraphics[height=4cm]{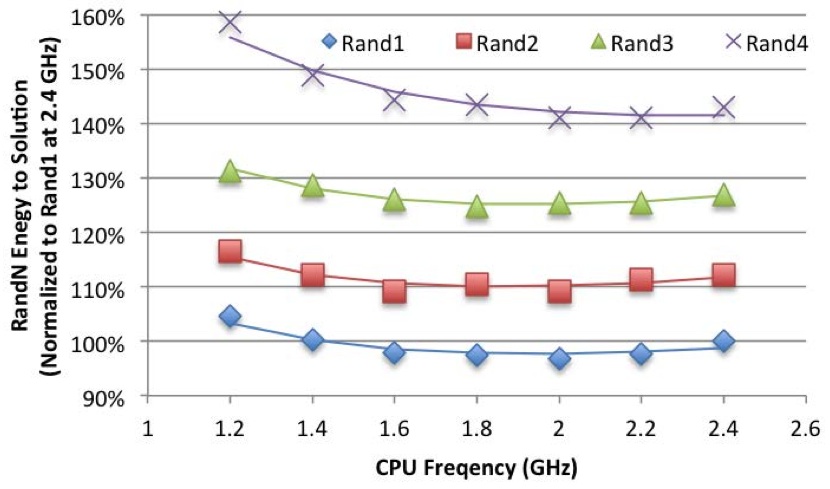}}
	\subfloat[][Senn \ea~\cite{DBLP:journals/ejasp/SennLJM05}]{\includegraphics{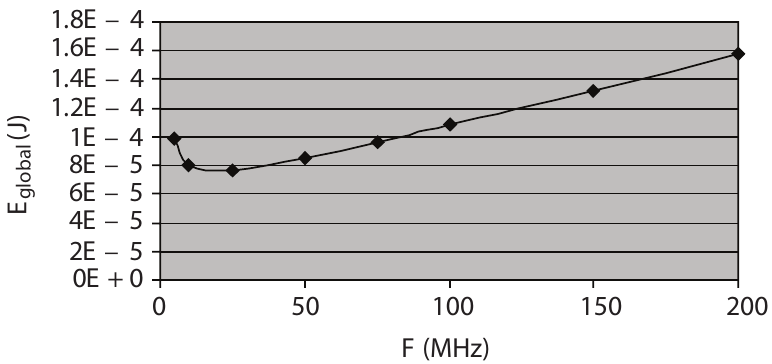}\label{fig:Senn}}
	\caption{Excerpts of energy/frequency measurements as found in the literature. Convex minimums are observable for the energy at a certain microprocessor clock frequency, depending on the microprocessor and architecture. In the sequel the behavior of this convex minimum is analyzed. All figures were originally published in the papers referenced in their respective captions.\label{fig:examples:convex}}
    \end{center}
\end{figure*}

There are some works that cover the energy/frequency convexity properties in a more analytical framework.
Figure~\ref{fig:examples:convex} shows excerpts of convex energy graphs provided by the cited works.
Yuki and Rajopadhye~\cite{yuki2013folklore} explored the particular case of energy consumption of high-performance computers in the context of compiler optimization and optimal frequency conditions of the microprocessor.
One of their conclusions is that for power-hungry systems the \emph{race-to-halt} energy optimization technique is more effective than \ac{DVFS}.
Hager~\ea~\cite{DBLP:journals/corr/abs-1208-2908}, on the other hand, showed that race-to-halt is not always the most effective strategy in a multi-core context with bandwidth-bound codes.
The authors studied the energy consumption of modern multi-core chips via simple machine models and showed how to minimize the energy consumption with respect to the number of cores, serial code performance, and clock frequency.
Austin and Wright~\cite{Austin:2014:MIM:2689710.2689716} examined the energy consumption of micro-benchmarks and applications on a Cray {\small CX}30 super computer system.
The authors developed a simple linear heuristic energy model.
They also stressed that the frequency/energy minimum is application-specific.
Cho and Chang~\cite{4167982} assessed the optimal frequency conditions for a microprocessor in conjunction with a memory.
Their resulting model is fairly complex; yet the authors show the feasibility of a microprocessor's optimal frequency conditions in conjecture with a memory system.
Cho and Melhelm~\cite{10.1109/TPDS.2009.41} produced a convex model derived from Amdahl's law and extended with the notion of energy.
The authors use a simplifying assumption for the representation of power and execution time.
They show via their model that there is a certain clock frequency range that yields both energy and speed improvements.
Similarly, Rizvandi~\ea~\cite{Rizvandi20111154} devised a convex model but, just as Cho and Melhelm, simplified representations of power and execution time were assumed.
Vasilaki~\cite{Vasilaki2015} showed experimental evidence for a convex energy curve in relation to the microprocessor's clock frequency for almost all individual instructions of the ARM Cortex A7.
No theoretical framework is provided by Vasilaki, however, to backup these findings analytically.

From an experimental perspective, Halimi~\ea~\cite{6682055} claim to save up to 39\,\% of energy, and Qiu~\ea~\cite{Qiu2012439} advertise an energy gain of 25\,\%, by adjusting the microprocessor's clock frequency via an experimental algorithm with predefined user or application constraints.
Although no theoretical framework was provided by the authors about the energy/frequency convexity, their algorithm is essentially chasing the convex minimum.
Senn~\ea~\cite{DBLP:journals/ejasp/SennLJM05} showed also convex energy/frequency curves, based on a simplified system model, for their {\small TI C55}, {\small C62}, {\small C64}, and {\small C67} platforms.

Applications of the work presented in this paper focuses on embedded systems, in contrast with Yuki and Rajopadhye's, Hager~\ea\ and Austin and Wright's work, which is dedicated to more powerful computer architectures.
The sensitivity of the parameters that constitute the energy consumption equation are also analyzed via both an analytical approach and via experimental data, the former fitted with data from the latter.
The convex energy model presented here is, in contrast with the mentioned works, more extensive, which allows for a more realistic modeling.
For example, temperature has not been a subject of interest and a sensitivity analysis of parameters has also not been carried out in any of the referenced works.

\section{Conclusion}
\label{sec:chapter:2b:conclusion}

In this paper we developed and analyzed the energy consumption equation of a microprocessor operating in a computer system with other components.
An analytical analysis, along with numerical simulation and measurement data, was used to study the behavior and sensitivity of its parameters.
It was shown through an analytical framework, measurements, and literature review that the energy consumption curve shows convex properties with regard to the clock frequency of the microprocessor.
The convex energy minimum is the point with a given clock frequency $\fo$ where the computer system consumes the minimum amount of energy while executing a code sequence.

The energy saving gained by running at the optimal clock frequency is a trade-off with the performance of the system, in terms of execution time.
For applications requiring human interaction, it has been shown by Seeker~\ea~\cite{DBLP:conf/iiswc/SeekerPLF14}, however, that the clock frequency can be scaled down considerably without affecting the user's experience.
More generally, this kind of energy savings can be obtained for code sequences where a limited slowdown can be tolerated and time is not critical.
For example, such slowdowns could be applied to code sequences, in multithreaded programs that are not on the critical path~\cite{DBLP:conf/green/HalimiPGJ14}.

The existence of the energy/frequency convexity property was further confirmed via experimental measurement traces of multimedia microprocessors commonly used for embedded system applications.
The main conclusions of the analysis are:
\begin{itemize}
 \item Energy/frequency convexity occurs always, but, to exploit the convex minimum, $\fo$ should be within the exploitable clock frequency window;
 \item The background power requirement ($\Pback$) is the parameter that influences the optimal frequency the most; the larger the background power demands, the larger the optimal clock frequency: when $\Pback$ equals $\Pcpu$, $\fo$ will be close to the maximum microprocessor clock frequency;
 \item An application's power profile ($\xi$) has a minimal effect on the optimal frequency, mostly because the variations in power profiles are fairly small in the experiments we ran, an average of 50\,MHz in $\fo$  between the power profile's extremities;
 \item The number of instructions of a code sequence has no influence on the optimal clock frequency, following the energy consumption model, but does scale the energy consumption linearly on the premise that $\xi$ has minimal effect at constant temperature;
 \item Application concurrency and clock cycle thieves ($\ck$) significantly affect the optimal frequency; the less clock cycles available to the applications, the larger the optimal clock frequency: on average for a 1\,GHz increase in $\ck$, $\fo$ increases by 2\,GHz;
 \item Microprocessor slack time ($\beta$), during off-chip operations, forces the optimal clock frequency down: 300\,MHz for 0 $ < \beta <$ 0.25 in the extreme case;
%
%
 \item The \emph{race-to-halt} strategy is justified only when the optimal clock frequency is larger than the microprocessor's maximum frequency.
\end{itemize}
Given that $\Pback$ has a large effect on the optimal frequency $\fo$, it was shown that a system with a $\Pback$ of the order of $\Pcpu$ and larger will have a $\fo$ likely outside the reach of the microprocessor's clock frequency range.
Thus chasing the optimal clock frequency $\fo$ is especially beneficial for low-power systems, such as for embedded applications, as their $\Pback$ is much smaller than what would be expected for high-performance computer systems.

%
%
%

\ifCLASSOPTIONcaptionsoff
  \newpage
\fi

\bibliographystyle{IEEEtran}
\bibliography{../../library/library}

\end{document}